\documentclass[aps,pra,notitlepage,superscriptaddress,floatfix,10pt,nofootinbib,twocolumn]{revtex4-1}
\usepackage[utf8]{inputenc}
\usepackage[english]{babel}
\usepackage[T1]{fontenc}
\usepackage{amsmath}

\usepackage{tikz}
\usepackage{lipsum}
\usepackage{graphicx}%
\usepackage{amsmath}
\usepackage{rotating}
\usepackage{extarrows}
\usepackage{color,calc,graphicx}
\usepackage{amsmath}
\usepackage{multirow}
\usepackage{mathrsfs}
\usepackage{bbm}
\usepackage{euscript}
\usepackage{amssymb}
\usepackage{extarrows}
\usepackage{bbm}
\usepackage[ruled]{algorithm2e}
\usepackage{graphicx}
\usepackage{epstopdf}
\usepackage{dcolumn}
\usepackage{bm}
\usepackage[all]{xy}
\usepackage{indentfirst}
\usepackage{amsmath}
\usepackage{multirow}
\usepackage{mathrsfs}
\usepackage{euscript}
\usepackage{amssymb}
\usepackage{appendix}
\usepackage{color,xcolor}

\newtheorem{theorem}{Theorem}

\newtheorem{proposition}{Proposition}
\begin{document}

\title{Multiparticle entanglement classification with ergotropic gap}

\author{Xue Yang}
\affiliation{School of Information Science and Technology, Southwest Jiaotong University, Chengdu 610031, China}
\affiliation{College of Computer Science and Cyber Security, Chengdu University of Technology, Chengdu, 610059, China}

\author{Yan-Han Yang}
\affiliation{School of Information Science and Technology, Southwest Jiaotong University, Chengdu 610031, China}

\author{Shao-Ming Fei}
\affiliation{School of Mathematical Sciences, Capital Normal University, Beijing 100048, China}
\affiliation{Max-Planck-Institute for Mathematics in the Sciences, 04103 Leipzig, Germany}

\author{Ming-Xing Luo}
\email{mxluo@swjtu.edu.cn}
\affiliation{School of Information Science and Technology, Southwest Jiaotong University, Chengdu 610031, China}

\begin{abstract}
The presence of quantum multipartite entanglement implies the existence of a thermodynamic quantity known as the ergotropic gap, which is defined as the difference between the maximal global and local extractable works from the system. We establish a direct relation between the geometric measure of entanglement and the ergotropic gaps. We show that all the marginal ergotropic gaps form a convex polytope for each class of quantum states that are equivalent under stochastic local operations and classical communication (SLOCC). We finally introduce the concept of multipartite ergotropic gap indicators and use them to present a refined criterion for classifying entanglement under SLOCC.
\end{abstract}

\maketitle

\section{Introduction}

Thermodynamics, a fundamental branch of physics, explores the relationships among heat, work, and energy in physical systems. An important facet of thermodynamics involves extracting work from isolated quantum systems through cyclic Hamiltonian processes \cite{Allahverdyan2004,Viguie2005,Skrzypczyk2014,Perarnau2015}. In quantum mechanics, the maximum work that can be extracted, known as ergotropy, is determined by the system's density matrix and Hamiltonian \cite{Allahverdyan2004, Vin(2016),Ciampini2017,Francica2017,Andolina2019,Monsel2020,Opatrny2021}. The maximal ergotropy under global and local cyclic Hamiltonian processes provides a new feature to characterize thermodynamic procedures. In particular, the ergotropic gap, which is the difference between the maximal global and local extractable works, has recently received widespread attention \cite{Mukherjee2016,Alimuddin2019,Alimuddin2020a,Alimuddin2020,Liu2021,Salvia2022,Seah2021,Tirone2021,Salvia2023,Yang2023}.

Multipartite quantum states serve as essential resources for various applications in quantum communication, quantum computing, and interferometry. One key problem  regarding these states is whether they can be categorized solely based on local information. The exploration of this marginal problem originated in the context of the well-known Pauli exclusion principle for fermions \cite{Coleman,Klyachko}. One common approach to categorizing entanglement is through stochastic local operations and classical communication (SLOCC) \cite{Horodecki2009}. For pure quantum states, leveraging single-particle information alone can serve as a means to detect multi-particle entanglement \cite{Walter2013}. Specifically, the spectral vectors of reduced densities of individual one-body systems collectively form an entanglement polytope for each entanglement class under SLOCC. The violation of these generalized polytope inequalities offers an effective method to detect multipartite entanglement locally. This classification of entanglement has also been realized experimentally \cite{Aguilar2015,Zhao2017}.

Work and quantum entanglement are fundamental resources in thermodynamics and quantum information theory, respectively. The theory of bipartite entanglement exhibits deep parallels with thermodynamics. \cite{Horodecki2002,Horodecki2009,detection2009}. While the importance of quantum entanglement is well-established in quantum information theory \cite{Horodecki2009,detection2009}, further exploration of the connections between work extraction and entanglement is warranted. Recent studies have revealed strong links between entanglement and ergotropic gaps \cite{Mukherjee2016,Alimuddin2019,Alimuddin2020,Puliyil2022}. Notably, the presence of quantum entanglement always leads to a non-zero ergotropic gap \cite{Mukherjee2016}. However, the direct relationship between entanglement classification under SLOCC and ergotropic gaps remains an open question that requires further investigation.

In this paper, we delve into the interplay between entanglement classification under SLOCC and ergotropic gaps. Our investigation centers on the marginal ergotropic gaps resulting from the partitioning of a single-qubit and the remaining qubits. We establish a direct link between these marginal ergotropic gaps and the geometric measure of entanglement. By leveraging polygonal inequalities \cite{Walter2013}, we show a crucial requirement for multi-qubit pure states: each marginal ergotropic gap must not exceed the sum of the others. Furthermore, we integrate these findings with the concept of entanglement polytopes to demonstrate that the vectors of marginal ergotropic gaps collectively form another polytope for each entanglement class. This ergotropic gap polytope holds physical significance in connection to the entanglement polytope and offers an additional criterion for identifying SLOCC multipartite entanglement classes. To distinguish between overlapping entanglement polytopes, such as those arising from generalized W states and GHZ states, we introduce a multipartite marginal ergotropic gap indicator. This indicator serves to identify SLOCC entanglement classes that cannot be validated by the entanglement polytopes \cite{Walter2013}.

\section{Ergotroic gap for qubit systems}

Consider a finite-dimensional system in the state $\varrho$ on Hilbert space $\mathcal{H}$ with a bare Hamiltonian $H$ at a given time period $\tau>0$.  The average extractable energy of the system is denoted as the expectation value of
\begin{eqnarray}
E(\varrho)={\rm Tr}(\varrho H).
\end{eqnarray}

According to the Schr\"{o}dinger dynamics, the unitary evolution $U(\tau)$ yields the final state to be $\varrho(\tau)=U(\tau)
\varrho U^{\dagger}(\tau)$. The work extracted from the system is then given by the difference between the initial and final energies:
\begin{eqnarray}
W_e(\varrho)={\rm Tr}(\varrho H) -{\rm Tr}(\varrho(\tau)H(\tau)).
\end{eqnarray}
For a specific Hamiltonian $H=\sum^d_{i=1}\epsilon_iE|\epsilon_i\rangle\langle\epsilon_i|$ with $\epsilon_1 \leq \cdots \leq \epsilon_{d}$, the maximal extractable work from the system in the initial state $\varrho$, known as ergotropy \cite{Allahverdyan2004}, is given by
\begin{eqnarray}
W_e(\varrho)\nonumber&=&{\rm Tr}(\varrho H)-\min_{U(\tau)}{\rm Tr}(U(\tau)\varrho U^{\dagger}(\tau)H)
\\
&=&{\rm Tr}(\varrho H) -{\rm Tr}(\varrho^p H),
\end{eqnarray}
where the passive state $\varrho^p$ is the minimum energetic state and is given by $\varrho^p=\sum^d_{i=1}\lambda_i|\epsilon_i\rangle\langle \epsilon_i|$, with $\lambda_1\geq\cdots \geq \lambda_{d}$.

For a bipartite $\varrho_{AB}$ on Hilbert space $\mathcal{H}_A\otimes \mathcal{H}_B$, the global bare Hamiltonian of joint system takes the form $H_{AB}=H_A\otimes \mathbb{I}_B+ \mathbb{I}_A\otimes H_B$, where $\mathbb{I}$ is the identity operator, the local Hamiltonian is $H_{X}=\sum^{d_X-1}_{j=0}\epsilon^{X}_jE|\epsilon^{X}_j\rangle\langle \epsilon^{X}_j|$ with eigenenergies $\epsilon^{X}_jE$ satisfying $\epsilon_j\leq \epsilon_{j+1}$, and $E$ denote the unit energy and $|\epsilon^{X}_j\rangle$ is the associated eigenstate. The maximal extractable work or the global ergotropy is then defined as
\begin{eqnarray}
W^G_e(\varrho_{AB})={\rm Tr}(\varrho_{AB}H_{AB})-{\rm Tr}(\varrho^{p}_{AB}H_{AB}),
\label{global0}
\end{eqnarray}
where $\varrho^p_{AB}$ is the corresponding passive state of $\varrho_{AB}$.  In contrast, the local ergotropy is defined for individual systems as
\begin{eqnarray}
&W^L_e(\varrho_{AB})=W^A_e(\varrho_{AB})+W^B_e(\varrho_{AB})
\nonumber
\\
&={\rm Tr}(\varrho_{AB}H_{AB})-{\rm Tr} (\varrho^p_{A}H_A)-{\rm Tr} (\varrho^p_{B}H_B).
\label{local0}
\end{eqnarray}
Both the global and local ergotropies allow defining the ergotropic gap as the extra gain given by \cite{Mukherjee2016}:
\begin{eqnarray}
&\Delta(\varrho_{AB})=W^G_e(\varrho_{AB})-W^L_e(\varrho_{AB})
\nonumber
\\
&={\rm Tr} (\varrho^p_{A}H_A)+{\rm Tr} (\varrho^p_{B}H_B)-{\rm Tr}(\varrho^p_{AB}H_{AB}).
\label{gap1}
\end{eqnarray}
This quantity characterizes the difference between the global and local maximum extractable works. In particular,  in the case of a bipartite pure state $\varrho_{AB}$, we have
\begin{eqnarray}
&\Delta(\varrho_{AB})={\rm Tr} (\varrho^p_{A}H_A)+{\rm Tr} (\varrho^p_{B}H_B),
\label{energypure}
\end{eqnarray}
which means that $\Delta(\varrho_{AB})$ represents the total extractable energy, derived from measuring the passive states of each subsystem A and B.

It has shown that the ergotropic gap can be applied to characterize the bipartite entanglement \cite{Mukherjee2016}. In what follows, we extend to feature multipartite entanglement.

In general, an $n$-qubit entangled pure state on Hilbert space $\mathcal{H}_1 \otimes \cdots \otimes \mathcal{H}_n$ can be written into
\begin{eqnarray}
|\Psi\rangle=\sum_{j=0}^n \sum_{s_j=0}^1 \alpha_{s_1 \cdots s_n}\left|s_1 \cdots s_n\right\rangle_{12 \cdots n},
\end{eqnarray}
where $\alpha_{s_1 \cdots s_n}$ are the coefficients satisfying the normalization condition of $\sum_{s_1 \cdots s_n}\left|\alpha_{s_1 \cdots s_n}\right|^2=1$. Denote the reduced density matrix of the qubit $i$ by $\varrho_i, i=1, \cdots, n$. For each $\varrho_i$, there are two eigenvalues $\left\{\lambda_{\min }^{(i)}, 1-\lambda_{\min }^{(i)}\right\}$ with $\lambda_{\min }^{(i)} \in\left[0, \frac{1}{2}\right]$.
Consider all the marginal ergotropic gaps for an $n$-qubit state with the bipartition $i$ and $\overline{i}=\{j\not=i,\forall j\}$, denoted as $\Delta_{i}$. Herein, each qubit $i$ is assumed to be governed by the Hamiltonian $H_i=E|1\rangle\langle 1|$ under local unitary operations for $1 \leq i\leq n$. The composite system is governed by the Hamiltonian $H=\sum^n_{i=1}H_i\otimes_{k\in\overline{i}} {\mathbb I}_{k}$, herein, $H_i\otimes_{k\in\overline{i}} {\mathbb I}_{k}={\mathbb I}_{1}\cdots {\mathbb I}_{i-1}\otimes {H}_{i}\otimes{\mathbb I}_{i+1}\cdots {\mathbb I}_{n}$. According to Eq.(7),  the marginal ergotropic gap (MEG) is then given by
\begin{eqnarray}
\Delta_{i}(|\Psi\rangle)=2\lambda^{(i)}_{min}E=2(1-\lambda^{(i)}_{max})E,
\label{puregap1}
\end{eqnarray}
which is related to the geometric measure of bipartite entanglement, i.e., the minimal eigenvalue of the state  \cite{Wei2003}. This reveals a remarkable correspondence between the thermodynamic quantity and the operational information for any $2\times 2^{n-1}$-dimensional isolated system.

\begin{figure}
\begin{center}
\resizebox{210pt}{120pt}{\includegraphics{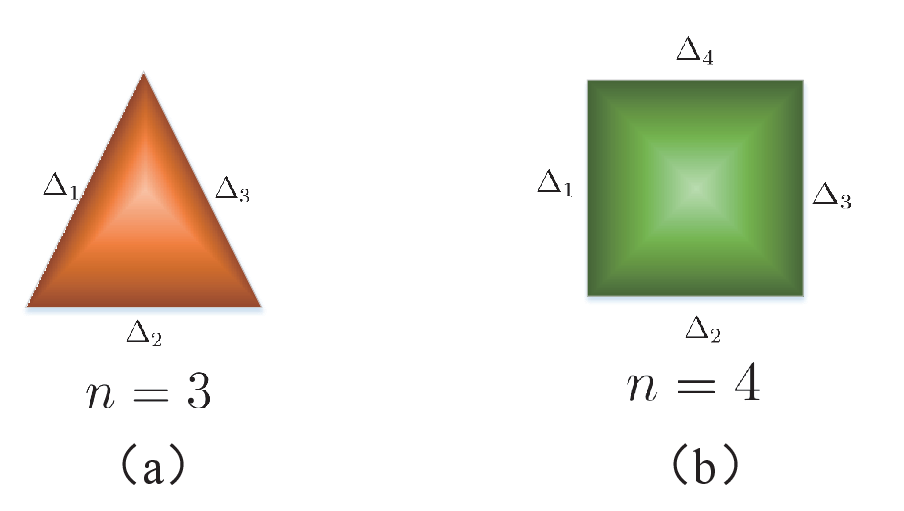}}
\end{center}
\caption{\small (Color online). Schematic polygon inequalities of MEG. (a) 3-qubit pure states; (b) 4-qubit pure state;  The length of each side represents correspondingly to the value of MEG.}
\label{polygontu}
\end{figure}

For any $n$-qubit pure state $|\Psi\rangle$ on Hilbert space $\mathcal{H}_{1}\otimes \cdots \otimes \mathcal{H}_n$, it has shown that the spectrum of the reduced density matrices satisfies the polygon inequalities \cite{Higuchi2003}:
\begin{eqnarray}
\lambda^{(i)}_{min}\leq \sum^n_{j\neq i,\forall j=1}\lambda^{(j)}_{min},
\label{polygon}
\end{eqnarray}
where $\lambda^{(i)}_{min}\in[0,\frac{1}{2}]$ is the smallest eigenvalue of the reduced density matrix $\varrho_i$ of the $i$-th qubit. Combined with  Eq.(\ref{puregap1}), the MEG satisfies the following polygon inequality
\begin{eqnarray}
\Delta_{i}(|\Psi\rangle)\leq \sum^n_{j\neq i,\forall j=1}\Delta_{j}(|\Psi\rangle),
\label{gappolygon}
\end{eqnarray}
where the local Hamiltonian is defined as $H_i=E|1\rangle\langle 1|$ for the $i$-th qubit. Remarkably, the inequality (\ref{gappolygon}) guarantee that the MEGs form a closed $n$-side energy polygon. This allows for a geometric representation of the inequality (\ref{gappolygon}) in terms of the MEGs as shown in Figure~1.

Let $\Delta^{total}(|\Psi\rangle)$ denote the total MEG for all possible bipartitions of subsystems, that is,
\begin{eqnarray}
\Delta^{total}(|\Psi\rangle)=\sum^n_{j=1}\Delta_{j}(|\Psi\rangle).
\end{eqnarray}
From the inequality (\ref{gappolygon}), the total work gap satisfies the following inequality as
\begin{eqnarray}
2\Delta_{j}(|\Psi\rangle)\leq \Delta^{total}(|\Psi\rangle), \forall j=1, \cdots, n.
\label{aa}
\end{eqnarray}
This provides an operational relationship of the MEGs of all bipartitions.

\section{Many-body entanglement classification}

The entanglement among quantum systems is highly relevant in thermodynamics, as the system with more entanglement may  have a higher ergotropic gap \cite{Alimuddin2019}. Our goal here is to explore many-body entanglement classification using the MEGs based on the quantum marginal problem related to the multipartite representability problem in quantum chemistry \cite{Walter2013}. Especially, consider an $n$-qubit quantum state $|\Psi\rangle$ on Hilbert space $\otimes_{i=1}^n\mathcal{H}_i$. It is entangled if it cannot be written as a product state of single qubit states \cite{Horodecki2009}. Given two states, they are equivalent under SLOCC if and only if they can be transformed into each other with local invertible operations. Define a characteristic vector of the MEG as
\begin{eqnarray}
\Lambda=(\Delta_{1}(|\Psi\rangle),\Delta_{2}(|\Psi\rangle),\cdots,\Delta_{n}(|\Psi\rangle)),
\label{puregapN}
\end{eqnarray}
where $\Delta_{i}(|\Psi\rangle)$ satisfies the inequality (\ref{gappolygon}). Denote ${\cal C}=G\cdot{}\rho$ as the SLOCC entanglement class containing $\rho$, where $G$ denotes a local invertible action such that $\varrho=G\cdot\rho$. From Eq.(\ref{gappolygon}) all the MEGs for any state in $\mathcal{C}$ form a convex polytope.

\begin{theorem}
For a given entangled pure state $|\Psi\rangle$ on Hilbert space $\otimes_{i=1}^n\mathcal{H}_i$, the MEG polytope of an entanglement class $\mathcal{C}=G\cdot|\Psi\rangle$  is given by the following convex hull
\begin{eqnarray}
\Delta_{\cal C}={\rm conv}\{(\Delta_{1},\Delta_{2},\cdots,\Delta_{n})\}.
\end{eqnarray}

\label{convex}
\end{theorem}

\emph{Proof.} Consider a system of $n$ qubits. Denote the marginal reduced density matrix of the qubit $i$ by $\varrho_i$. For each $\varrho_i$, let  $\lambda^{(i)}_{\max }$ be the largest  eigenvalue of each one-qubit density matrix $\rho_i$. It turns out that the vectors $\vec{\lambda}=(\lambda^{(1)}_{\max}, \lambda^{(2)}_{\max},\cdots,\lambda^{(n)}_{\max})$  associated with all pure states $|\Psi\rangle$ in the closure of an orbit under SLOCC transformations form a polytope \cite{Walter2013}. As a result, Theorem \ref{convex} can be proven using \cite[Theorem 2]{Walter2013} and the relation (\ref{puregap1}). $\Box$

Eq.(\ref{puregap1}) establishes a relation between the marginal ergotropic gap, denoted as $\Delta_{i}$, and the maximal eigenvalue of each one-qubit density matrix. Additionally, the maximal eigenvalues, which represent the vertices of entanglement polytopes, can be computed from covariants within a finitely generated algebra \cite{Walter2013}. Consequently, the vertices of the marginal ergotropic gap polytopes, i.e., $\Delta_{i}$, are determinable. Furthermore, each $\Delta_{i}$ can be directly detected in experiments by measuring only the single-particle energy of the corresponding passive for each qubit. This work focuses on presenting a thermodynamic method for identifying multipartite entanglement.

Theorem \ref{convex} reveals new meanings of entanglement in terms of physical thermodynamics and provides a criterion for entanglement classification under SLOCC. It has been shown that two transformable multipartite states under SLOCC \cite{Chen(2011)}, where  the Schmit tenser rank has been introduced, of which the crucial idea is to find the minimal decomposition on the product basis. However, even for qubit case, the evaluation of tensor rank is an NP-hard problem. Instead, our goal here is to present a computationally efficient method. We only use the marginal information of mutlipartite qubit states as $\Delta_{i}$ for $i=1,2,\cdots, n$. Marginal ergotropic gap polytopes provide a simple way for identifying global feature of multi-particle entanglement. If the set of ergotropic gaps of the reduced density matrices of a given pure state $|\psi\rangle$ does not fall into the   polytope $\Delta_{\cal C}$ of marginal ergotropic gaps, the given state cannot belong to the entanglement class $\cal C$: $\Lambda_\psi \notin \Delta_{\mathcal{C}} \Rightarrow |\psi\rangle \notin \mathcal{C}$. This procedure requires linear-time complexity.

Our approach to detecting entanglement in Theorem 1 remains applicable for featuring some noisy states. Specially, consider a minimum bound $1-\varepsilon$ on the purity $\operatorname{Tr}\rho^2$ of a quantum state $\rho$, which implies a fidelity $\langle\psi|\rho| \psi\rangle \geq 1-\varepsilon$ with a pure state $|\psi\rangle$. In this case, all the local eigenvalues of $\rho$ are different from those of $|\psi\rangle$ by an amount $\delta(\varepsilon)$. For an $n$-qubit system, the total deviation is approximately $\delta(\varepsilon) \approx N \varepsilon / 2$ \cite{Walter2013}. Thus, if the measured marginal ergotropic gap vector $\Lambda$ of the state $\rho$ are at a distance of at least $2\delta(\varepsilon)$ from the marginal ergotropic gap polytope $\Delta_{\mathcal{C}}$ defined in Theorem 1, the prepared state $\rho$ exhibits a high fidelity with some pure state that is more entangled than the class $\mathcal{C}$.

We  explain the present approach with an example. Suppose that $\rho$ is an experimentally prepared four-qubit state with the purity $1-\varepsilon=0.9$. Then by the above there exists a pure state $|\psi\rangle$ with the fidelity $\langle\psi|\rho| \psi\rangle \geq 0.9$, according to ref.\cite{Walter2013}, the 1-norm difference of the maximum eigenvalues satisfy that
\begin{eqnarray}
\sum_k |\lambda_{\max }^{(k)}(\rho)-\lambda_{\max }^{(k)}(|\psi\rangle)|\leq 0.21.
\label{noise}
\end{eqnarray}
It has also been demonstrated that if the inequality $\sum^4_{k=1}\lambda_{\max }^{(k)}<3$ holds, then a given state $|\psi\rangle$ does not belong to W-type entanglement. By combining the aforementioned inequality with Eq. (9), we obtain
\begin{eqnarray}
\Delta_1(|\psi\rangle)+\Delta_2(|\psi\rangle)+\Delta_3(|\psi\rangle)+\Delta_4(|\psi\rangle)>2,
\end{eqnarray}
which implies that $|\psi\rangle$ is not entangled in terms of the W-type, see more details shown in Example 2. This means that it is  sufficient by using Eqs.(\ref{noise}) and (\ref{puregap1}) to verify that the single-particle eigenvalues of the experimentally prepared state $\rho$ satisfy the relation
\begin{eqnarray}
\Delta_{1}(\rho)+\Delta_{2}(\rho)+\Delta_{3}(\rho)+\Delta_{4}(\rho)>2.42.
\end{eqnarray}
This provides a method to characterize the noisy quantum states.

\textit{Example 1}. For a three-qubit pure state, there exists a one-to-one correspondence between six entanglement classes and  entanglement polytopes  \cite{Walter2013,Dur}. Note that the entanglement polytopes are given by the vectors $\vec{\lambda}=\left(\lambda^{(1)}_{\max }, \lambda^{(2)}_{\max }, \lambda^{(3)}_{\max }\right)$. The vertices of entanglement polytopes can be computed from so-called covariants that do not vanish identically on the orbit \cite{Walter2013}.
Combining this fact with Eq.(\ref{puregap1}), Theorem \ref{convex} implies that the marginal ergotropic gap polytopes of all pure states consist of a convex hull of 5 vertices. One vertex, $(\Delta_{1},\Delta_{2},\Delta_{3})=(0,0,0)$, corresponds to the product states. Three vertices, (0,1,1), (1,0,1) and (1,1,0), are for three kinds of biseparable states. And the last  (1,1,1)  corresponds to the maximally entangled GHZ state $|GHZ\rangle=\frac{1}{\sqrt{2}}(|000\rangle+|111\rangle)$ \cite{GHZ}, as shown in Figure~2. Here, the lower pyramid represents the generalized W states given by:
\begin{eqnarray}
|W\rangle=a_1|001\rangle+a_1|010\rangle+a_2|100\rangle
\label{Wstate}
\end{eqnarray}
 with $\sum_i|a_i|^2=1$. Note the vector $(\lambda^{(1)}_{max}, \lambda^{(2)}_{max}, \lambda^{(3)}_{max})$, corresponding to the collection of local maximal eigenvalues, is contained in the W-type entanglement polytope  \cite{Walter2013}. This implies that
\begin{eqnarray}
\lambda^{(1)}_{max}+\lambda^{(2)}_{max}+\lambda^{(3)}_{max}\geq2.
\label{GHZ1}
\end{eqnarray}
By combining the inequality (\ref{GHZ1}) and Eq. (\ref{puregap1}), the MEG vector satisfies the following inequality
\begin{eqnarray}
\Delta_1+\Delta_2+\Delta_3\leq2E.
\label{GHZ}
\end{eqnarray}
Moreover, the entire polytope is for the generalized GHZ states:
\begin{eqnarray}
|GHZ\rangle=\cos\theta|000\rangle+\sin\theta|111\rangle
\label{GHZstate}
\end{eqnarray}
with $\theta\in (0, \frac{\pi}{4})$.

\begin{figure}
\begin{center}
\resizebox{200pt}{90pt}{\includegraphics{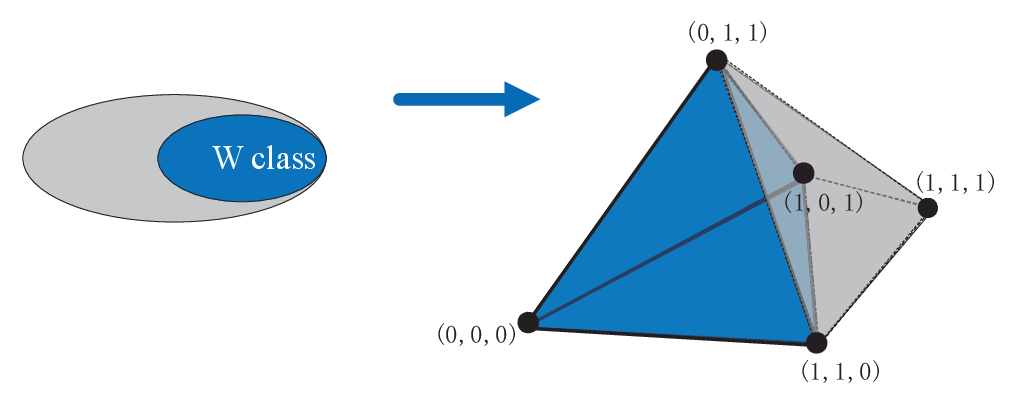}}
\end{center}
\caption{\small (Color online). The MEG polytopes of three-qubit pure states.}
\label{polytope}
\end{figure}

Equation (\ref{GHZ}) provides a new classification scheme under the SLOCC for three-qubit pure states in terms of the MEG. The violation of the inequality (\ref{GHZ}) witnesses GHZ-type entanglement. Howbeit, a MEG vector may fail to fully distinguish different entanglement classes. This can be further resolved by using the following work indicators as
\begin{eqnarray}
\eta(|\Psi\rangle)=\min_{j} \left\{\Delta_{j}(|\Psi\rangle)-\sum^n_{k\neq j,k=1}{\Delta}_{k}(|\Psi\rangle)\right\}.
\label{indicatorMEG}
\end{eqnarray}
The quantity $\eta(|\Psi\rangle)$ characterizes the genuine tripartite entanglement \cite{Bai3,Luo2022}, a tripartite pure state $|\psi\rangle \in \mathcal{H}_1 \otimes \mathcal{H}_2 \otimes \mathcal{H}_3$ is called biseparable if it can be written as $|\psi\rangle=$ $\left|\psi_i\right\rangle \otimes\left|\psi_{jk}\right\rangle$.  If a tripartite state is not biseparable then it is called genuinely tripartite  entangled, i.e., the genuine tripartite entanglement cannot be written as a product state in terms of any bipartition.

\begin{proposition}
\label{Threequbit}
Suppose the bar Hamiltonian of each qubit is $H_j=E|1\rangle\langle1|$. A three-qubit pure state $|\Psi\rangle$ is genuine tripartite entangled if $\eta(|\Psi\rangle)\neq 0$.

\end{proposition}

\textit{Proof}. Assume that the quantum state $|\Psi\rangle$ on Hilbert space $\mathcal{H}_1\otimes \mathcal{H}_2\otimes \mathcal{H}_3$ is not genuine tripartite entangled, i.e., it allows the following decomposition as
\begin{eqnarray}
&&|\Psi_{ijk}\rangle=|\varphi_{i}\rangle\otimes|\varphi_{jk}\rangle,
\label{bipartite0}
\\
&&|\Psi_{ijk}\rangle=|\phi_{i}\rangle\otimes|\phi_{j}\rangle\otimes|\phi_{k}\rangle,
\label{bipartite1}
\end{eqnarray}
where $i\neq j\neq k\in\{1,2,3\}$, $|\varphi_{i}\rangle$ and $|\phi_j\rangle$ denote the states of the respective system $i$ and $j$, and $|\varphi_{jk}\rangle$ denotes the state of the joint system $jk$. For the case (\ref{bipartite0}), from (\ref{puregap1}), we obtain $\Delta_i=0$ and $\Delta_j=\Delta_k$ from the pure state $|\varphi\rangle_{jk}$. Moreover, it is easy to show that $\Delta_j=\Delta_i+\Delta_k$ for all $i\neq j\neq k\in\{1,2,3\}$, where $\Delta_{j}$ denote the ergotropic gap under the bipartition $j$ and $ik$.  This implies that $\eta (|\Psi\rangle)=0$ for any $j\in \{1,2,3\}$. Similar result holds for the case (\ref{bipartite1}). $\Box$

As a complement to Theorem 1, the multipartite ergotropic gap indicator $\eta$ provides a finer criterion to  classify multi-qubit pure states under the SLOCC. The present MEG-based criterion on SLOCC classification can identify states better than the conclusion from ref.\cite{Walter2013}, see the following examples.

\emph{Example 2}. Consider a generalized $n$-qubit W state on Hilbert space $\otimes_{i=1}^n\mathcal{H}_i$ given by
\begin{eqnarray}
|W_n\rangle&=&\sum^n_{i=1}\sqrt{a_i}|\vec{1}_i\rangle_{1\cdots n},
\label{Wn}
\end{eqnarray}
where $\vec{1}_i$ denotes all zeros except for the $i$-th component which is $1$, $a_i$ satisfy $\sum^n_{i=1}a_i=1$ and $a_1\geq\cdots \geq a_{n}$. Suppose that the bar Hamiltonian of each qubit $i$ is $H_i=E|1\rangle\langle 1|$ for $1\leq i\leq n$. The reduced density matrix of the subsystem $i$ is given by
\begin{eqnarray}
\varrho_{i}=(1-a_i)|0\rangle\langle0|+a_i|1\rangle\langle1|.
\end{eqnarray}
 According to Eq.(\ref{puregap1}),  if there exists $ a_i>\frac{1}{2}$, we obtain that $\Delta_{i}=2(1-a_i)E$ and $\Delta_{j}=2a_jE$, whereas  we have $\Delta_{i}=2a_iE$ for any $a_i<\frac{1}{2}$.  This further implies that
 \begin{eqnarray}
\Delta^{total}(|W_n\rangle)=
\left\{
\begin{aligned}
  &4(1-a_i)E<2E,    && \exists i, a_i>\frac{1}{2},
  \\
  &2E,    && \forall i, a_i<\frac{1}{2}.
\end{aligned}
\label{Wn0}
\right.
\end{eqnarray}
Hence, we obtain that $\Delta^{total}(|W_n\rangle)\leq 2E$ for any $n$-qubit W state. This shows that the generalized W state with $a_i<1/2$ for any $i$ lies on the facet of $\Delta^{total}(|W_n\rangle)=2E$.

\textit{Example 3}. Consider a generalized $n$-qubit GHZ state on Hilbert space $\otimes_{i=1}^n\mathcal{H}_i$ given by
\begin{eqnarray}
|GHZ_n\rangle&=\cos\theta|0\rangle^{\otimes n}+\sin\theta|1\rangle^{\otimes n})_{1\cdots n},
\label{GHZg}
\end{eqnarray}
where $\theta\in(0,\frac{\pi}{4}]$. The bar Hamiltonian of each qubit $i$ is $H_i=E|1\rangle\langle 1|$ for $1\leq i\leq n$. The reduced density matrix of the subsystem $i$ can be written
\begin{eqnarray}
\varrho_{i}=\cos^2\theta|0\rangle\langle0|+\sin^2\theta|1\rangle\langle1|.
\end{eqnarray}According to Eq.(\ref{puregap1}), the MEG under the bipartition $i$ and $\overline{i}$ is given by $\Delta_{i}(|GHZ_n\rangle)=2\sin^2\theta E$. This implies that
\begin{eqnarray}
\Delta^{total}(|GHZ_n\rangle)=2n\sin^2\theta E\leq n E,
\label{GHZn1}
\end{eqnarray}
where the maximum gap is obtained from the maximally entangled $n$-qubit GHZ state, i.e., $\theta=\frac{\pi}{4}$. Moreover, we have
\begin{eqnarray}
\Delta^{total}(|GHZ_n\rangle)>2E
\label{GHZn2}
\end{eqnarray}
for $\theta>\arcsin \sqrt{1/n}$. This means the generalized GHZ state with $\theta>\arcsin\sqrt{1/n}$ does not belong to the facet determined by the generalized $n$-qubit W state in Example 2. This privies a way to distinguish two entanglement classes by using the facet of $\Delta^{total}=2E$.

However, for the case of $\theta<\arcsin \sqrt{1/n}$, we have
\begin{eqnarray}
\Delta^{total}(|GHZ_n\rangle)<2E.
\label{GHZn}
\end{eqnarray}
According to Eqs.(\ref{GHZn}) and (\ref{Wn0}), the presence of overlapping regions fails to distinguish W-type entanglement from GHZ-type entanglement. The indicator $\eta$ may be used to certify the entanglement as
\begin{eqnarray}
&&\eta(|W_n\rangle)=0,\exists i, a_i>\frac{1}{2},
\\
&&\eta(|GHZ_n\rangle)=-2(n-2)\sin^2\theta E<0,~~ \forall i.
\end{eqnarray}
Moreover, for the case of $\theta=\arcsin \sqrt{1/n}$, we have
\begin{eqnarray}
\eta(|GHZ_n\rangle)=-\frac{2(n-2)E}{n}.
\label{GHs}
\end{eqnarray}
Meanwhile, for the generalized W state we obtain that
\begin{eqnarray}
&&\eta(|W_n\rangle)=4a_iE-2E,\forall i, a_i<\frac{1}{2}.
\end{eqnarray}
This implies $\eta(|GHZ_n\rangle)\not=\eta(|W_n\rangle)$ for any one $a_i$ satisfying $a_i\not=\frac{1}{n}$. So, we have distinguished the generalised GHZ state from all the almost generalized W states except for the maximally entangled W state beyond recent result \cite{Walter2013}.

\emph{Example 4}. Consider an $n$-qubit Dicke state with $l$ excitations \cite{Stockton2003} given by
\begin{eqnarray}
|D^{(l)}_n\rangle_{12\cdots n}=\frac{1}{\sqrt{\binom{n}{l}}}\sum_{\textsf{g}\in S_n}\textsf{g}(|0\rangle^{\otimes n-l}|1\rangle^{\otimes l})
\end{eqnarray}
for $1\leq l\leq n-1$, where summation is over all possible permutations $\textsf{g}\in S_n$ of the product states with $n-l$ number of $|0\rangle$ and one $|1\rangle$, $S_n$ denotes the permutation group on $n$ items, and $\binom{n}{l}$ denote the combination number choosing $l$ items from $n$ items. Herein, each qubit $i$ is governed by the Hamiltonian $H_i=E|1\rangle\langle 1|$. By the symmetry, the reduced density matrix of any subsystem $i$ is given by
\begin{eqnarray}
\varrho_i=\frac{n-l}{n}|0\rangle\langle0|+\frac{l}{n}|1\rangle\langle1|.
\end{eqnarray}
According to Eq.(\ref{puregap1}), it is easy to check that
\begin{eqnarray}
\Delta_i(|D^{(l)}_n\rangle)
=\left\{
\begin{aligned}
  &\frac{2lE}{n},    &&  l\leq\frac{n}{2},
  \\
  &\frac{2(n-l)E}{n},    &&  l>\frac{n}{2}.
\end{aligned}
\right.
\label{}
\end{eqnarray}
This implies a general polytope facet as
\begin{eqnarray}
\Delta^{total}(|D^{(l)}_n\rangle)
=\left\{
\begin{aligned}
  &2lE,    &&  l\leq\frac{n}{2},
  \\
  &2(n-l)E,    &&  l>\frac{n}{2}.
\end{aligned}
\right.
\label{DickeN}
\end{eqnarray}
Especially, $|D^{(l)}_n\rangle$ is reduced to the $n$-qubit W state when $l=1$ and $n\geq3$. This implies $\Delta^{total}(|W\rangle)=2E$. A further evaluation shows for any generalized Dicke states that
\begin{eqnarray}
|\hat{D}^{(l)}_n\rangle=\sum_{\textsf{g}\in S_n}\alpha_{\textsf{g}}\textsf{g}(|0\rangle^{\otimes n-l}|1\rangle^{\otimes l})_{12\cdots n},
\end{eqnarray}
where $\alpha_{\textsf{g}}$ depending on the permutation $\textsf{g}\in S_n$ satisfy $\sum^{n \choose l}_{\textsf{g}}\alpha_{\textsf{g}}^2=1$. From the symmetry $\Delta_i(|\hat{D}^{(l)}_n\rangle)$ takes the maximum value for the case of $\alpha_{\textsf{g}}=1/\sqrt{\binom{n}{l}}$ for each $\alpha_{\textsf{g}}$. As a result, we obtain
\begin{eqnarray}
\Delta^{total}(|\hat{D}^{(l)}_n\rangle)\leq
\left\{
\begin{aligned}
  &2lE,    &&  l\leq\frac{n}{2};
  \\
  &2(n-l)E,    &&  l>\frac{n}{2}.
\end{aligned}
\right.
\end{eqnarray}
These show that the generalized Dicke states are under the facet of $\Delta^{total}(|\hat{D}^{(l)}_n\rangle)< 2lE$, while $n$-qubit Dicke states $|D^{(l)}_n\rangle$ is on the facet of $\Delta^{total}(|D^{(l)}_n\rangle)=2lE$. It provides a way to witness Dicke states with different excitations that are inequivalent under the SLOCC, that is, the $n$-qubit Dicke states with different excitations lies on different facets.

\section{Conclusion}

In this paper, by defining the difference between the maximal global and local extractable works, we show the ergotropic gap plays a significant role in certifying quantum entanglement. This has given the profound analogies between thermodynamic quantities and entanglement. The present criterion for SLOCC classification based on MEG can distinguish quantum states more accurately compared to previous results from ref.\cite{Walter2013}. For an arbitrary many-body isolated system, it shows that a nonzero ergotropic gap is a necessary and sufficient condition to guarantee the entanglement \cite{Alimuddin2019}. The bipartite ergotropic gap can further capture the figure of genuineness in multipartite entanglement, which may lead to new genuine multipartite entanglement measures from bipartite ergotropic gaps. Furthermore, it is still unknown whether the set of MEG for multipartite mixed states forms a convex polytope.

\section*{Acknowledgments}

This work was supported by the National Natural Science Foundation of China (Nos.62172341, 12204386, 12075159, and 12171044), Sichuan Natural Science Foundation (Nos.2023NSFSC0447, 24NSFSC5421), Beijing Natural Science Foundation (No.Z190005), Interdisciplinary Research of Southwest Jiaotong University China Interdisciplinary Research of Southwest Jiaotong University China (No.2682022KJ004), and the Academician Innovation Platform of Hainan Province.


\end{document}